\documentclass{egpubl}
\usepackage{eurova2020}

 \WsPaper           

 \electronicVersion 

\ifpdf \usepackage[pdftex]{graphicx} \pdfcompresslevel=9
\else \usepackage[dvips]{graphicx} \fi

\PrintedOrElectronic

\usepackage{t1enc,dfadobe}

\usepackage{egweblnk}
\usepackage{cite}

\usepackage{caption}
\usepackage[table]{xcolor}
\usepackage{xspace}
\definecolor{c1}{rgb}{0,0.3,1}
\definecolor{c2}{rgb}{1,0,0.0}
\definecolor{c3}{rgb}{0.16, 0.5, 0.0}
\definecolor{c4}{rgb}{1, 0.73, 0}
\definecolor{c5}{rgb}{0.2, 0.64, 0.38}

\usepackage{marginnote}

\newcommand{\newtechniquename}{\textit{SpatialRugs}\xspace}
\newcommand{\motionrugs}{\textit{MotionRugs}\xspace}
\usepackage[normalem]{ulem}
\usepackage{microtype}
\title[SpatialRugs]%
      {SpatialRugs: Enhancing Spatial Awareness of Movement\\ in Dense Pixel  Visualizations\vspace{-1.5\baselineskip}}
\author[Buchm\"uller et al.]
{\parbox{\textwidth}{\centering Juri F. Buchm\"uller$^{1}$ and
        Udo Schlegel$^{1}$ and
        Eren Cakmak$^{1}$ and
        Evanthia Dimara$^{1}$ and 
        Daniel A. Keim$^{1}$
        \vspace{-1.5\baselineskip}
        }
        \\
{\parbox{\textwidth}{\centering $^1$University of Konstanz, Data Analysis and Visualization Group, Germany
       }
}
}

\graphicspath{{figures/}{pictures/}{images/}{./}}
\begin{document}

\teaser{
\vspace{-3em}
 \includegraphics[width=.95\linewidth]{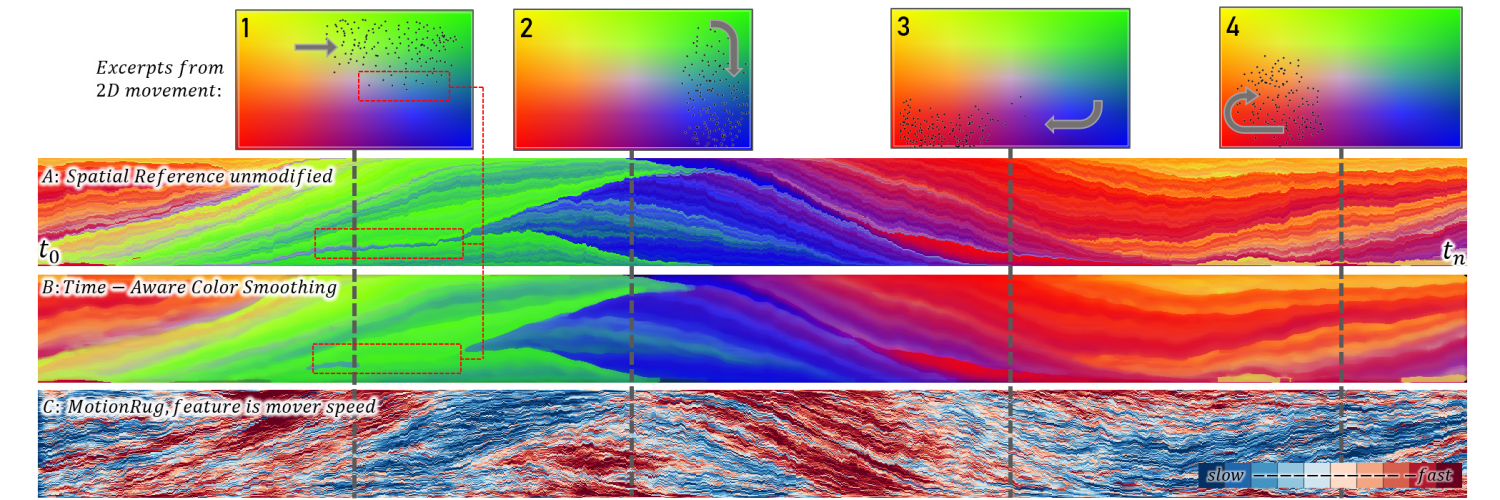}
 \centering

  \caption{\newtechniquename (A+B) and \motionrugs (C), all with the same underlying dataset of 151 fish moving in a tank for about 90 seconds. 
  Excerpts 1-4 show static snippets of the fish turning 
  from the upper right over the lower right to the lower left. 
  Part A shows unmodified \newtechniquename, where colors can be related to spatial positions (compare colors to Parts 1-4). 
  Part B shows color-smoothed \newtechniquename that mitigate distorted patterns (outlined in red boxes). 
  Part C shows mover speed encoded in the colors instead of the position. 
  In conjunction, \newtechniquename  and \motionrugs  can be applied to relate space to features (e.g., in which areas of A movers are fast or slow as indicated in C.)}
\label{fig:teaser}
}

\maketitle
\begin{abstract}
   Compact visual summaries of spatio-temporal movement data often strive to express accurate positions of movers. We present \newtechniquename, a technique to enhance the spatial awareness of movements in dense pixel visualizations.  \newtechniquename apply 2D colormaps to visualize location mapped to a juxtaposed display. We explore the effect of various colormaps discussing perceptual limitations and introduce a custom color-smoothing method to mitigate distorted patterns of collective movement behavior.

\end{abstract}
\vspace{-10em}

\vspace{-4em}

\section{Introduction}
\begin{figure*}[ht!b]
  \centering
  \includegraphics[width=1\linewidth]{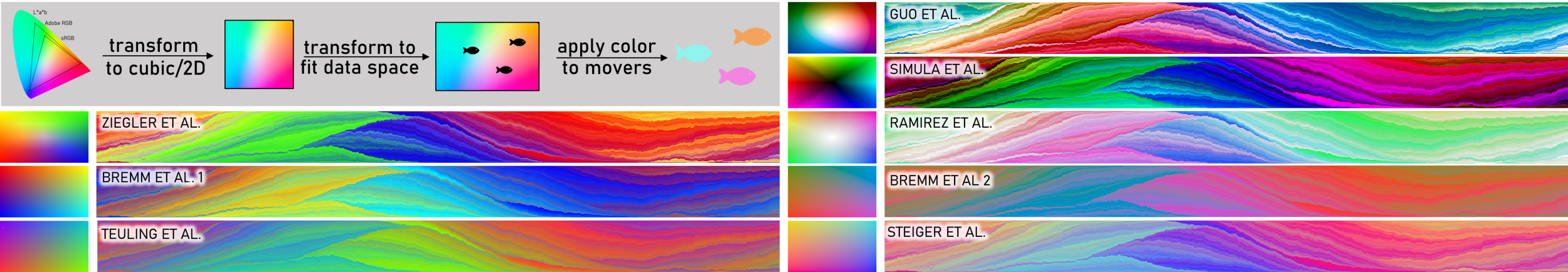}
  \caption{\label{fig:rugcol}%
    Upper left: In \newtechniquename, a color space is transformed into a 2D cubic, then adapted to the extent of the moving area. 
    A position is then encoded using the correspondent color from the color space. 
    Below: Application examples of different colormaps ~\cite{bernard2015survey} to the demo dataset containing 151 movers showing collective behavior. Left of each visualization, we see the underlying transformed 2D color space.
          }
          \vspace{-1.5\baselineskip}
\end{figure*}
Visualizations of movement data face challenges in scalability towards time and amount of displayed movers.
Especially, uncovering spatio-temporal patterns in collective movement behavior is challenging 
due to large numbers of entities moving similarly over long periods.
To overcome such scalability issues, the \motionrugs technique has been proposed that displays movers in a static, compact fashion~\cite{buchmuller2018motionrugs}. 
In \motionrugs (Figure~\ref{fig:teaser}~C), each pixel represents one mover, while the X-axis denotes time and the Y-axis represents a 1D spatial aggregation of all movers. Color can encode any numeric feature of interest, e.g., the speed of the entities. To illustrate, Figure~\ref{fig:teaser}~C shows the speed of the movers; several trends of slowing down (red) and speeding up (blue) can be visible at a glance, while the curvature reveals the spatial dynamics of the collective behavior (e.g., change in orientation and positions of the group).
Despite the space-efficient benefits of \motionrugs, users cannot relate movers to their original locations due to the spatial linearization, as is possible with many other techniques including simple static plotting or animation of trajectories~\cite{Andrienko2013Movement}. This is a major drawback, as spatial context can be important when analyzing movements. For example, to be able to explain movers' behavior, locations of food sources for animals or points of interest for human movers can reveal critical contextual information.

In this paper, we combine the space-efficiency benefit of \motionrugs with the space-awareness advantages of other advanced techniques for trajectory visualization~\cite{andrienko2013space,hurter2009fromdady,tominski2012stacking}.
We propose \newtechniquename (see Figure~\ref{fig:teaser} A and B), a technique 
that builds on the application of 2D color maps to dense pixel visualizations, 
transforming colors to express the spatial positions of movers.
We further refine \newtechniquename with a time-aware color correction to mitigate perceptual issues arising from color space transformations (see Figure~\ref{fig:teaser} B). We compare the results to a naive gaussian-based color smoothing approach and discuss suitable color spaces.
\section{MotionRugs \& Collective Movement Visualization} 

Visual analysis of movement capitalizes on human perception to process summaries of complex movement data and uncover 
patterns over time and space \cite{Andrienko2013Movement}. 
Andrienko et al.~\cite{andrienko2013space} provide an example of spatial abstraction for collective movement, transforming physical mover positions to relative ones regarding a group-centered reference point. 
\motionrugs ~\cite{buchmuller2018motionrugs} further reduce the space of the moving entities from
a 2-D to an 1-D representation, 
ideally still reflecting
the physical distances between the movers as accurate as possible. 
To create the 1-D order from a set of 2-D positions of movers at a given moment in time, 
spatial linearization strategies are used such as space-filling curves or the traversal of spatial index structures~\cite{lu1993spatial}.  
In \motionrugs~\cite{buchmuller2018motionrugs}, every mover in one frame is represented by a single pixel and
colored after the feature a user is interested in 
(e.g.,  speed in Fig.~\ref{fig:teaser}~C). 
The process is repeated for each time frame, 
ordering the slices on the x-axis by time. 
The result is a static, dense pixel display~\cite{keim2001visual}, 
showing the feature development of the movers over time. 
Alternative dense representations for geometrical relation exist, 
such as ParaGlide~\cite{bergner2013paraglide} for parameter exploration of animal behavior models, 
or Cui et al.~\cite{cui2014let}, showing remarkable results for dynamic graphs. 

In contrast, \motionrugs~\cite{buchmuller2018motionrugs} are primarily used to retain a quick overview of the spatial dynamics reflected in inherent curved spatial dynamics patterns, which can as well be employed to detect trends in features and feature distributions.
%
%
However, unlike other sophisticated techniques for trajectory visualization \cite{andrienko2013space,hurter2009fromdady,tominski2012stacking}, \motionrugs lacks spatial awareness, as it does not depict the accurate spatial locations of the movers.
While \motionrugs capture changes in space and mover orientation over time, it is unable to show 
\textit{where} entities are moving to specifically. 
This limitation is critical for many use cases where analysts needs to be aware of the region the entities are moving in. 
To enhance spatial awareness, while preserving the spatial efficiency of \motionrugs, we propose an alternative approach. 
Below, we propose \newtechniquename, a technique that reintroduces spatial positions into \motionrugs, eliminating the necessity for tedious analyses (e.g., clutter-prone static trajectory plots or time-consuming animations).
\section{Retaining spatial readability with \newtechniquename}
\begin{figure*}[h!tb]
  \centering
  \includegraphics[width=.95\linewidth]{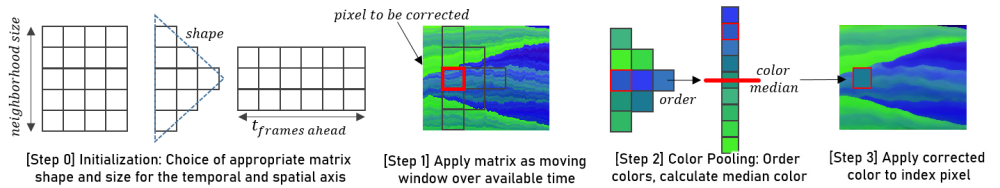}
  \caption{\label{fig:correction}%
    Process of the pooling-based color correction. First, an appropriate matrix shape and size needs to be determined according to the specific use case. One matrix dimension determines the size of the neighborhood to be included, the other determines the amount of time ahead to be considered for the correction. The matrix is then shifted over every pixel in every time step. In each step the colors of the matrix cells are ordered by euclidean distance in the RGB-space. The median color is then used as corrected color and applied to the original pixel.}
          \vspace{-1.5\baselineskip}
\end{figure*}
\newtechniquename is a compact visualization technique for collective movement data that enhances spatial awareness by projecting a 2D-color space into the 1D-linearizaton of \motionrugs.
\newtechniquename apply a color scale method to the original movement space. 
Fig.~\ref{fig:rugcol} (upper left corner) demonstrates our approach:
(i) We transform the color space to a 2-D cubic representation in order to serve as a base for the second step.
(ii) We transform the 2-D color space to cover the maximum extent of the spatial dimensions used by the mover dataset.
(iii) We assign the 2-D position of a mover to the corresponding color of the transformed color map. 
Spatial positions are now represented by color, 
which can be used in conjunction with pixel-based visualizations of movement, such as \motionrugs~\cite{buchmuller2018motionrugs},
to encode mover locations.  
With the colormap reference, users are able to identify the spatial distribution of entities at a given time. 
Fig.~\ref{fig:teaser} shows that the movers come from the upper right corner (green, first excerpt), 
take a right turn towards the lower right (blue, second excerpt), move through the lower middle of the represented space in purple to the lower (red, third excerpt) and finally middle left in orange color tones (fourth excerpt). 
The resulting patterns allow to perceive the movers' spatial distribution, while viewers can also estimate how the movers progress within the color zones. For example, between excerpts 1 and 2, just a few movers start to move towards the blue until everyone follows. This behavior is shown as cone-shaped transition from green to blue. Consequently, the color mapping enables to compactly see patterns over long periods of time, also relating the spatial development to the feature development by comparing the excerpts (e.g. by relating Fig.~\ref{fig:teaser} A and C).
\vspace{-.5em}
\section{Color Space Considerations}\label{sec:colorspaceconsideration}
Color space mappings have been applied to represent spatial or temporal relations before. Northern Lights Maps~\cite{janetzko2009northern} map spatio-temporal properties of movers to an RGB-color scale. PhenoVis~\cite{leite2016phenovis} presents color-coded normalized stacked bar charts to allow comparative analysis over longer time spans. MotionExplorer ~\cite{bernard2017visual} employs a projection-based view displaying human motions in a 2D color-coding to highlight temporal patterns.
Similarly, \newtechniquename apply a 2D color space mapping to allow users to map colors of data points in an abstract visualization to their real spatial positions. As reflected in the 2D color map task assessment ER1-3 by Bernard et al. ~\cite{bernard2015survey}, a viewer should be able to  distinguish different locations by comparing their color representations accurately \textbf{(I)}. Also, the colors should be distributed as evenly as possible over the available space \textbf{(II)}, and finally, our approach should allow for two or more locations to be compared with each other \textbf{(III)}.

Yet, the visible spectrum and derived standard color spaces, e.g., CIELAB, HSV or sRGB are mostly organized in three dimensions and not necessarily of a symmetrical shape. So, the transformation to a regular 2-D form as required by \newtechniquename is challenging. Also, even disregarding errors introduced by color space transformation, color perception is individually different in viewers~\cite{dasgupta2018effect}, resulting in different abilities to identify fine-grained colormaps. Thus, a sensible choice of color space is critical for the effectiveness of \newtechniquename.
Many related approaches have employed 2D colormaps in specific and generic use cases. 
In an extensive survey, Bernard et al.~\cite{bernard2015survey} investigate the capabilities of 22 different 2-D color maps with respect to analytical tasks and perceptual properties.\\
\noindent\emph{Task assessment:} Fig.~\ref{fig:rugcol} shows a comparison of the color maps taken from Bernard et al.\cite{bernard2015survey} generated with the data described in Fig.~\ref{fig:teaser}.
According to the task assessment table of Bernard et al., colormaps provided by Bremm et al.~\cite{bremm2011assisted}, Ramirez et al.~\cite{ramirez2012self}, Steiger et al.~\cite{steiger2014visual} and Teuling et al.~\cite{teuling2011bivariate} would be best suitable given our defined tasks \textbf{I-III}. Yet, the task-based recommendations \cite{bernard2015survey} do not regard the perceptibility of visual structures \textit{within} the visualization space. As retaining these structures is important to our approach, we turn to the quality assessment measures \cite{bernard2015survey}.

\noindent\emph{Quality assessment:} The JND measure describes the ``Just Noticeably Different Colors''~\cite{bernard2015survey}, indicating how well a colormap exploits a color space. Here, the colormaps by Simula and Alhoniemi~\cite{simula1999som} and Guo et al~\cite{guo2005multivariate} perform well, but iterate over black or white in the center of the color map. 
Such color maps with low black- or white distance score work well only in conjunction with backgrounds of the opposite color ~\cite{bernard2015survey}. As \newtechniquename pack pixels densely and do not feature intermediate spaces between the data points, using color maps with black or white color ranges could interfere with the perceived brightness and saturation of the surrounding colors, making the color map difficult for our case. The next best color maps according to the JND feature are Cube Diagonal Cut B-C-Y-R ~\cite{bremm2011assisted} and the Four Corners R-B-G-Y color map ~\cite{ziegler2007visual}. 

\noindent\emph{Transformation assessment:} Besides applying an appropriate color map, the visual outcome of \newtechniquename is also determined by the amount of applied transformation to the color space. Changing the ratio of an original color space in one axis affects the color discriminability along the same axis. This holds even if the ratio is changed in both axes. In both directions (either shrinking or enlarging the color space), color discriminability suffers, since either there will be less space to represent all colors a color space can provide, or the same colors are stretched over a larger space than the color space can cover. Yet, since color perception is not necessarily linear, these effects can only be measured in perceptual studies. While we acknowledge these effects, we expect that our technique is still applicable to aspect rations of up to 16:9.

\noindent\emph{Distribution assessment:} Movement distributions play a crictical role for the visual outcome of \newtechniquename. If only a few movements take place in a narrow color range, the visual discriminability of rest of the movements is reduced, but outliers could still be seen well. To solve this issue, an adaptive coloring approach based on the movement distribution would be conceivable in future work.
\vspace{-.5em}

\section{Pooling-based Time Aware Color Smoothing}

Transition areas in the color space frequently appear as perceptual distortions (outliers), for example, Fig.~\ref{fig:teaser} (excerpt 1, outlined in red) shows most movers in the green quadrant and only a few in the transition area to the blue quadrant. 
This results in a salient blue line (outlined in the red box), 
where the perceived color distances appear larger than 
the actual distances of the blueish movers to the rest of the green group.
Such artifacts can mislead viewers to think that the few entities are further into the blue space as they are in reality. 
To mitigate such perceptual distortions, we propose a time-aware color smoothing technique. 
Our method includes the mover distribution 
of the current and subsequent steps to determine the color correction.
If entities close to each other are located in different color areas, their respective color is corrected towards the majority. 
\begin{figure}[t!]
  \centering
  \includegraphics[width=.95\linewidth]{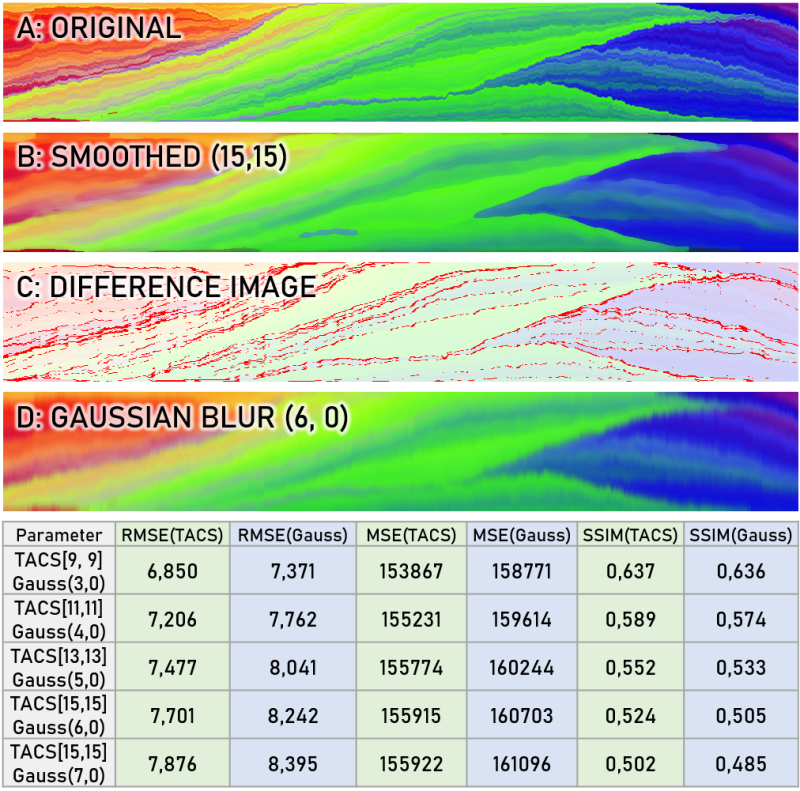}
  \caption{\label{fig:smoothing-comparison}
    Comparing an unmodified SpatialRug(A) to a smoothed one (B) and gaussian blur (D). C provides a difference image between A and B and highlights the areas our smoothing focuses on in red. The table shows quantitative assessment results for time-aware color smoothing(TACS) versus standad gaussian blur(Gauss). 
          }
          \vspace{-2em} 
\end{figure}

Our method (seen in Fig.~\ref{fig:correction}) consists of three steps (color collection, pooling, and adaption) repeated for every pixel. 
During the initialization phase (Fig.~\ref{fig:correction}, Step 0), 
users adjust the pooling matrix, selecting three parameters:
\emph{neighborhood size}, \emph{time frames} ahead, 
and \emph{step size}. 
The neighborhood size describes the spatial region around the focused pixel in vertical axis.
The time frames ahead incorporates the spatial movement into the future to smooth in horizontal direction.
Lastly, the step size offers a way to reduce the neighborhood into the future to steer the importance of the developing spatial region. 
Please refer to our supplementary material for alternative parameter variations. The code for the color smoothing is publicly available as Python notebook~\cite{schlegeltacs}.
In Step 1, we apply the user-defined pooling matrix around the target pixel and collect the colors of included pixels. 
In step 2, the collected pixels are ordered with a stable sorting algorithm (e.g., mergesort) on the RGB values.
Outlier pixel color will be sorted to both ends of the list, while more similar colors move to the mid.
In Step 3, after the sorting, the median of the array yields the color average of the collected pixels and the index pixel is corrected to the median.

\section{Results: assessing visual outcomes}
We next elaborate on preliminary results on the effectiveness of \newtechniquename , discuss color scale choice and smoothing method.\\ 
\noindent \emph{Color scale}: In section~\ref{sec:colorspaceconsideration}, we proposed an initial set of color maps to explore using the work of Bernard et al.~\cite{bernard2015survey} and defined the requirements \textbf{I-III}.
We further narrow down the selection of well applicable colormaps by visually investigating color space properties (see Figure~\ref{fig:rugcol}).
First, derived colors should be well distinguishable to relate them to an accurate spatial location, satisfying requirement \textbf{I}. The color maps provided by Bremm et al. 2~\cite{bremm2011assisted}, Steiger et al.\cite{steiger2014visual} and Teuling et al.\cite{teuling2011bivariate} are clearly inferior to their competitors for this property. Second, requirement~\textbf{II} states that colors should be distributed as evenly as possible. Here, the color map provided by Simula et al.~\cite{simula1999som} introduces a black/dark area between neighboring colors in the corners, impacting the perceptual continuity. The color regions by Ramirez et al.~\cite{ramirez2012self} and Bremm et al. 1~\cite{bremm2011assisted} are also not linearly distributed.
This leaves the colormaps by Ziegler et al.~\cite{ziegler2007visual} and Guo et al.~\cite{guo2005multivariate} as candidates. Ziegler et al. anchors four distinctive colors, amongst them three primary colors, to the corners of the color space creating a semantic notion of spatial orientation resembling the intuitive natural division of four cardinal directions. Guo et al. extend the color space radially around a white center. Both approaches scale well to different aspect rations, satisfying requirement ~\textbf{III}. With the approach of Guo et al., an additional center area can be encoded using a white color. Yet, this could interfere perceptually, if an additional feature should be encoded as modification of the color brightness, which only works if no black or white components are present. To leave this possibility open, Ziegler et al.'s approach is more suitable. In conclusion, we expect the color maps by Ziegler et al. and Guo et al. to fulfill our requirements. The choice between the two is use case-dependent.\\
\noindent \emph{Color smoothing}: 
%
The time-aware smoothing aims to mitigate the effects of neighboring colors (outlined in red Fig~\ref{fig:smoothing-comparison}~A) by including the temporal color distribution.
In Fig.~\ref{fig:smoothing-comparison} A and B, we see that the methods reduce visible outliers while retaining the temporal structures.
The difference image between (A) and (B) (see Fig.~\ref{fig:smoothing-comparison}~(C)) provides preliminary evidence for the value of the applied smoothing method as it only affects the color transition areas, leaving the visual patterns still crisp and visible.
In contrast, the Gaussian blur (D) creates a fuzzy impression, aggravating accurate interpretation of colors at a given point by blurring visual structures.
A quantitative assessment of our color-smoothing (table in Fig.~\ref{fig:smoothing-comparison}) shows results of applied quality measures by measuring the distance to the original, unsmoothed image. These measures include the root mean squared error (RMSE)~\cite{image_quality_metrics}, the mean squared error (MSE)~\cite{image_quality_metrics} and the structural similarity index~\cite{ssim} (SSIM). We compare our time-aware color smoothing (TACS) to a standard gaussian smoothing (Gauss).
Similar reference area parameters are chosen to enable a better comparison of the smoothing methods.
Lower RMSE and MSE values indicate better results, whereas a higher value for SSIM indicates better similarity between original and smoothed image. 
The results indicate that our pooling method outperforms the Gaussian blur even for small sigmas and large window sizes.
\vspace{-.5em}

\section{Conclusion and Future Work}

We demonstrated an approach to encode collective movement behavior for spatial awareness
within a static, compact visualization. 
\newtechniquename uses color mapping to allow users to perceive spatial relations through space-efficient designs.
The intended use of \newtechniquename is in conjunction with other pixel-based visualizations of movement datasets showing other features the user is interested in, enabling him to relate space and feature developments (compare SpatialRug and MotionRug in Figure~\ref{fig:teaser}).
We compared several color spaces identifying advantages and disadvantages. Further comparisons of color spaces and results for the color smoothing on several color maps can be found in the supplemental material. 
We further discussed perceptual challenges derived by color scales, 
where movements appear more distant than in their physical space. 
To mitigate distortion effects, we proposed a time-aware color smoothing approach, which we illustrated 
in some examples and provided preliminary quality metrics. 
We expect that our approach can be applied to non-spatial 2D point distributions as well, for example to projections of dynamic datasets.

Despite its promising potential, \newtechniquename come with several shortcomings.
Spatial distance may introduce errors when users try to relate a color to its precise position, 
while individual differences in color perception might affect clarity of the derived patterns. 
All these aspects need to be evaluated, while guidelines for the correct parameterization of our technique have to be explored. 
In future work, we intend to quantify viewer's perception of our technique and choice of color spaces. 
Also, the perceptual implications of our color correction process have to be tested thoroughly. 
Instead of using a single color map, we anticipate that \newtechniquename would benefit from an adaptive color map approach adjusted to the specific movement distributions and user task. 
Finally, we would like to support better detection of movements in semantically interesting regions by allowing users to place color anchors interactively according to semantic objects or areas.

\bibliographystyle{eg-alpha-doi}  
\bibliography{egbibsample}        



\end{document}